\documentclass[a4paper]{article}
\usepackage{epsfig}
\usepackage{graphicx}

\begin{document}
\begin{titlepage}


\vfill
\begin{center}

{\bf\LARGE
\centerline{Identification of backgrounds in the}
\centerline{EDELWEISS-I dark matter search experiment}
}
\vfill

{\large The EDELWEISS Collaboration:} \\
S.~Fiorucci$^{1,*}$,
A.~Benoit$^{2}$,
L.~Berg\'e$^{3}$,
J.~Bl\"umer$^{4,5}$,
A.~Broniatowski$^{3}$,
B.~Censier$^{3}$,
A.~Chantelauze$^{5}$,
M.~Chapellier$^{7}$,
G.~Chardin$^{1}$,
S.~Collin$^{3}$,
X.~Defay$^{3}$,
M.~De~J\'esus$^{6}$,
H.~Deschamps$^{1}$,
P.~Di~Stefano$^{6}$,
Y.~Dolgorouky$^{3}$,
L.~Dumoulin$^{3}$,
K.~Eitel$^{5}$,
M.~Fesquet$^{1}$,
J.~Gascon$^{6}$,
G.~Gerbier$^{1}$,
C.~Goldbach$^{8}$,
M.~Gros$^{1}$,
M.~Horn$^{5}$,
A.~Juillard$^{3}$,
R.~Lemrani$^{1}$,
A.~de~Lesquen$^{1}$,
A.~Lubashevskiy$^{9}$,
M.~Luca$^{6}$,
S.~Marnieros$^{3}$,
L.~Mosca$^{1}$,
X.-F.~Navick$^{1}$,
G.~Nollez$^{8}$,
E.~Olivieri$^{3}$,
P.~Pari$^{7}$,
V.~Sanglard$^{6}$,
L.~Schoeffel$^{1}$,
F.~Schwamm$^{1}$,
M.~Stern$^{6}$,
E.~Yakushev$^{9}$
\end{center}

\vfill

{\scriptsize\noindent
$^{1}$CEA, Centre d'\'Etudes Nucl\'eaires de Saclay, DSM/DAPNIA, 91191 Gif-sur-Yvette Cedex, France\\
$^{2}$Centre de Recherche sur les Tr\`es Basses Temp\'eratures, SPM-CNRS, BP 166, 38042 Grenoble, France\\
$^{3}$Centre de Spectroscopie Nucl\'eaire et de Spectroscopie de Masse, UMR8609 IN2P3-CNRS, Univ. Paris Sud, b\^at~108, 91405 Orsay Campus, France\\
$^{4}$Institut f\"ur Experimentelle Kernphysik, Universit\"at Karlsruhe (TH), Gaedestr.~1, 76128~Karlsruhe, Germany\\
$^{5}$Forschungszentrum Karlsruhe, Institut f\"ur Kernphysik, Postfach 3640, 76021 Karlsruhe, Germany\\
$^{6}$Institut de Physique Nucl\'eaire de Lyon, Universit\'e de Lyon (Universit\'e Claude Bernard Lyon~1) et IN2P3-CNRS, 4 rue Enrico Fermi, 69622 Villeurbanne, France\\
$^{7}$CEA, Centre d'\'Etudes Nucl\'eaires de Saclay, DSM/DRECAM, 91191 Gif-sur-Yvette Cedex, France\\
$^{8}$Institut d'Astrophysique de Paris, UMR7095 CNRS - Universit\'e Pierre et Marie Curie, 98 bis Bd Arago, 75014 Paris, France\\
$^{9}$Laboratory of Nuclear Problems, JINR, Joliot-Curie 6, 141980 Dubna, Moscow Region, Russian Federation
}

\vfill

\begin{center}

{\bf Abstract}

\end{center}

This paper presents our interpretation and understanding of the
different backgrounds in the EDELWEISS-I data sets. We analyze in
detail the several populations observed, which include gammas,
alphas, neutrons, thermal sensor events and surface events, and try
to combine all data sets to provide a coherent picture of the nature
and localization of the background sources. In light of this
interpretation, we draw conclusions regarding the background
suppression scheme for the EDELWEISS-II phase.

\vfill

{\bf PACS classification codes:}
95.35.+d, 
14.80.Ly, 
98.80.Es, 
29.40.Wk. 

\vfill

{\scriptsize\noindent
$^{*}$Present address: Brown University, Department of Physics, Providence, RI 02912, USA
}

\end{titlepage}

\section{Introduction}
\label{par:intro} EDELWEISS-I is the first phase of an experiment
aiming at the direct detection of WIMPs (Weakly Interacting Massive
Particles) which could constitute the dark matter halo of our
Galaxy. It uses cryogenic detectors able to measure simultaneously
the heat and ionization components of the energy deposit induced by
the elastic scattering of a WIMP off a target nucleus (see
e.g.~\cite{Gaitsk} for a review).

The final analysis of the EDELWEISS-I data~\cite{San05} covers a
total fiducial exposure of 62 kg.d. The limits on the neutralino
scattering cross-section were obtained from the observation of 40
nuclear recoil candidates with recoil energies between 15 and 200
keV, of which 3 are between 30 and 100 keV. The limits were obtained
without the subtraction of any background, although the presence of
a coincidence between two detectors and the study of charge
collection distributions suggested that at least some of these
events are due to a neutron background and surface interactions of
electrons. In the preparation of the second phase of the experiment,
it was necessary to perform a thorough investigation of all events
in EDELWEISS-I before the nuclear recoil selection, in order to
better assess their origin and devise means to remove as many
background components as possible for EDELWEISS-II.

After a short description of the EDELWEISS-I experiment and of the
active background rejection capabilities of heat/ionization
cryogenic bolometers, this paper presents the evaluations of the
backgrounds related to different origins : gamma radiation, alpha
particles and other surface events, and neutrons. While the energy
range of interest for WIMP detection is limited to  below 100~keV,
high energy gamma and alpha lines can prove very useful to
understand the backgrounds in the range relevant to WIMP
interactions.

\section{EDELWEISS-I}

\subsection{General setup}
\label{par:setup}

EDELWEISS is located in the Laboratoire Souterrain de Modane (LSM)
which provides a $\sim$4800~m.w.e. rock shielding against cosmic
muons, reducing the vertical muon rate to 4.5 /day/m$^2$. The
320 g germanium detectors are operated at a very low temperature
(typically 17~mK) within a dilution cryostat. This cryostat, made
mostly of ultrapure copper, is further protected from external
radioactivity by 10~cm Cu and 15~cm Pb shields~\cite{Bel96}. Pure
nitrogen gas is circulated inside this shielding to reduce radon
accumulation. An external 30~cm layer of paraffin protects the
experiment from neutrons created in the rock. Inside the cryostat,
the detectors are shielded from the  radioactivity in components
of the electronics by 7~cm of roman lead \footnote{The
archeological lead comes from an antique roman ship. The wreck was
investigated during four campaigns (1984-86) supervised by M. L'Hour
of the Direction des Recherches Arch\'eologiques
Sous-Marines~\cite{Lhour}.}. The cold electronic components (in
total, nine FETs and a dozen of resistors and capacitances) are kept
as far away as possible from the detectors, and the rest of the
acquisition chain remains outside the cryostat. Fig. 1 presents a
schematic cut view of the cryostat inside its Cu and Pb shields.

The detectors themselves are encased into 1~mm thick ultrapure Cu
casings, and held in place with the help of teflon balls and three
small Cu springs (see the inset in Fig. 1). The thermal sensor
consists of a small (1 mm$^3$) Neutron Transmutation Doped (NTD) germanium
crystal glued directly onto each crystal. The thermal
coupling to the cryostat is assured by ultrasonic bonding of
several micrometric gold wires on gold pads. Electric connections
are assured by the same kind of gold wires linking the electrodes to
contact tracks on the copper casing. The wires going up to the 4~K
pre-amplifier level are low radioactivity coaxial cables. Further
details can be found in~\cite{martineau}.

\subsection{Active background rejection capabilities}
\label{par:rejcap}

It is impossible to completely shield the detectors from external
radiation. Most of the radiation reaching the detectors is in the
form of gamma rays. In EDELWEISS-I, the rate of gamma interactions
exceeds the one expected from WIMPs by at least a factor 10$^5$. In
order to address this, the EDELWEISS detector technology offers
means to actively discriminate between electron recoils caused by
photons and electrons, and nuclear recoils caused by neutrons or
WIMPs. Indeed, for the same interaction energy, nuclear recoils have
an ionization efficiency on average three to four times less than
electron recoils, depending on the energy~\cite{Quench}. By
measuring simultaneously a heat signal and an ionization signal and
considering the ratio of both parameters, it is possible to reject
more than 99.9~\% of the gamma interactions while keeping a 90~\%
efficiency for nuclear recoils down to an energy of
15~keV~\cite{San05}. The value of this threshold depends on the
experimental resolutions, which have been as low as 1.0~keV on the
ionization channel and 0.3~keV on the heat channel.

The ionization measurement is made possible by two aluminum
electrodes (thickness 60-100~nm) deposited onto top and bottom surfaces of the
germanium detector. By applying a moderate voltage between the
electrodes, charge carriers created by an interaction can be
collected and give rise to a signal. The selected voltage is 4~V,
which is high enough to efficiently collect charges but not too high
in order to preserve the discrimination capabilities. Furthermore,
one of the collection electrodes is actually separated into a center
part and an outer guard ring (see Fig. 2). This allows to
define a fiducial volume inside the detector where a reliable charge
collection is expected, as opposed to the lateral sides where
electric field lines can escape the crystal before they reach the
electrodes, leading to an incomplete charge collection. The current
NTD bolometer technology has been "upgraded" with the addition of a
$\sim$60~nm layer of amorphous Ge (GGA detector type) or Si (GSA detector type) just below the electrodes (see Fig. 2),
which essentially gets rid of the majority of surface
interactions~\cite{Shu00,Ben02}.

\section{Gamma background}
\label{par:gamma}

\subsection{Radioactivity measurements and material selection}
\label{par:g-meas} More than 99.5~\% of the interaction rate in the
EDELWEISS-I detectors is due to gamma events. The 15~cm thick
lead shield stops most of the gammas from outside the experimental
setup and the observed gamma background comes from the inside. To
select low activity materials a separate low-background counting
facility was built: a n-type coaxial High Purity Ge diode of
210~cm$^3$ operated at 77 K with archeological lead shielding. This
dedicated HPGe diode was not set up at the time of construction of
the shielding and of the cryostat. Thus, only a partial material selection
could be performed before the experiment;
the copper used for the shielding had to be
measured ``a posteriori''. The HPGe diode was intensively used for
internal material selection close to the detectors. Table 1 displays
the measured radioactivities of some components of the EDELWEISS-I
set-up, most of them situated in the immediate vicinity of the
detectors: copper detector holders, copper springs, teflon pads and
coaxial cables.

\subsection{Data sets and high-energy data reconstruction}

The present analysis is based on two data sets. The first is the run
labeled 2003p in Ref.~\cite{San05}, representing 39.4 kg.d of data
(total volume of the three detectors) taken in very stable
conditions, with similar performances of the three detectors in
terms of energy threshold and recoil energy resolution. For the WIMP
search, the amplifier gains, ADC bit ranges and channel
sensitivities are optimized for the low-energy WIMP signal, with the
consequence that ionization signals above 1 MeV saturate the
digitizers. Saturation of the heat signal occurs at~350 to 600 keV,
depending on the detector. This is far above the expected range for
WIMP signals, but it affects the identification of radioactive
backgrounds where gamma lines up to ~2.6 MeV can prove useful.

For this reason, a dedicated run, labeled 2003h, has been recorded
in the same experimental conditions as the run 2003p, but with all
amplifier gains reduced by a factor 10. The total exposure for this
run is 9.3~kg.d in the total volume of the three detectors.

Comparing the data from the runs 2003p and 2003h, it was
confirmed that the saturated signals of the run 2003p could be
corrected and their amplitudes could be calibrated reliably. This is
possible because, as described in Ref.~\cite{San05}, extensive
samples of the signal time profiles associated to each event are
stored onto disk. However, the filtering applied by the acquisition
system distorts the shape of saturated pulses. Therefore new
templates were built by filtering ideal events truncated at
different levels of saturation. The template used for a given event
is the one giving the best fit. It is obtained by varying the
saturation level of the template and minimizing the corresponding
$\chi^2$.

This allowed us to reconstruct gamma and alpha events with recoil
energies as high as ~7 MeV, and a full width resolution of about
3~\% at 2.6 MeV (ionization channel). This reconstruction method
suffers however from a lack of sensitivity in the case of weakly
saturated pulses as the procedure is comparatively more sensitive to
the baseline level of the events which has a direct impact on the
saturation level at a given energy. Conservatively, as discussed
in~\cite{FioTh}, the method is not used below 1 MeV for the
ionization channel, and below 350-600~keV for the heat channel,
depending on the detector.

\subsection{Experimental gamma background spectrum}

The energy calibration of ionization signals was performed
using $^{57}$Co (122 keV), $^{137}$Cs (662 keV) and $^{60}$Co (1173
and 1332 keV) sources. The full width resolution was about 2.5\% and
the response of the ionization channel was linear up to 1332 keV.

Fig. 3-a shows the ionization spectrum corresponding to the total
volume of the three detectors for the two data sets 2003h (9.3
kg.day) and 2003p (39.4 kg.day). In these spectra, the 2003p data
above 1 MeV have been reconstructed as described in the previous
section. As we are concerned with the gamma 
spectrum, the alpha events, which appear with low quenching values
essentially below 2~MeV ionization energy, have been eliminated by a cut
on the quenching factor $ Q (Q < 0.5)$, see 
below section 4. 
The most noticeable feature is a Compton backscattering
bump around 200~keV. Lines originating from U-Th series are visible,
notably the $^{208}$Tl line at 2614 keV with a full width resolution
of about 3\%. A peak of $^{40}$K at 1461 keV is also present.

\subsection{Background simulations}
\label{par:g-simu}

Monte Carlo simulations are performed under GEANT3~\cite{GEANT}
using the geometry given in Fig. 1 and the measured activities or
limits of Table 1. Several sources of gamma background are
successively studied and comparisons with the observed spectrum are
given in Fig. 3-b to 3-d.

\subsubsection{Copper detector holders and thermal shields}
The most massive materials in the immediate vicinity of the
detectors are detector holders, cryostat structure and thermal
shields. They are made of about 20 kg of ultra-pure copper (OFHC copper,
purity $>99.99\%$). The upper limit on their U/Th content is 0.1
ppb (Table 1). After machining, the copper holders were brought
underground only a few months before the start of the experiment.
Therefore, the cosmogenic activation of $^{60}$Co (half-life 5.3 y)
at the surface has to be considered. The experimental limit on
$^{60}$Co is compatible with an equilibrium concentration of about 1
mBq/kg~\cite{Heusser}. The U-Th and $^{60}$Co contributions,
assuming activities equal to the limits, are compared in Fig. 3-b
with the experimental spectrum. The simulated continuum is too low
by one order of magnitude, with much more pronounced lines than
actually observed.

\subsubsection{Radon}
In the lead-copper shield, near the detectors, about 20 liters of
air are trapped (see Fig. 1). Although continuously flushed with
pure nitrogen emanating from the liquid nitrogen dewar, this volume
is a potential source of radon contamination. No radon concentration
measurement has been made and the decay of $^{220}$Rn is simulated
assuming an activity of 10 Bq/m$^3$, the mean value measured in the
LSM cavity. As can be seen on Fig. 3-c, the simulated continuum is
again much too low, and the predicted lines are not observed. Even
with an unrealistic high concentration, the decay of radon can't
account for the observed spectrum.

\subsubsection{Copper shield}
The inner 10 cm thick copper shield (Fig.~1), is made of about 1 ton
of copper bricks (electrolytic copper, 99.9\% purity),
purchased and brought underground in the early nineties. Measured
``a posteriori'', this copper shows a small but measurable
contamination in U-Th series (Table 1). The cosmogenic $^{60}$Co is,
as expected, not detected after about two half-lives of decay
underground. The simulation shows that the U-Th content of this
copper shield accounts for most of the features of the background
energy spectrum (see Fig.3-d). Gammas originating from the bulk of
this copper shield go through several centimeters of material before
they reach the detectors, and, as a consequence, lines aren't very
pronounced, Compton diffusion being the dominant energy dissipation
process above $\sim$ 150 keV. The Compton 
backscattering bump around 200~keV is reproduced though the lower amplitude in the simulation indicates that an
other source of high energy gamma rays might be present.
The rate of the $^{208}$Tl line at
2614~keV (1.1$\pm$0.2 counts/kg/d)is reproduced within 30\% by the
simulation, a satisfying agreement given the uncertainty of 60\% on
the measured thorium concentration in copper.

\subsubsection{$^{40}$K}
A peak of $^{40}$K at 1461 keV (1.6$\pm$0.5 counts/kg/d) is
present in the energy spectrum. Table 1 shows that the highest
$^{40}$K activity is found in the wires to the detectors (teflon
sheath), but the involved mass is very low (few grams) and the simulated rate
is two orders of magnitude lower than the observed one. The same
holds for $^{40}$K contaminations of Cu springs (1.3 g for one detector) and teflon balls (0.4 g for one detector) in
the detector holders. The present measurement of the copper of the
gamma shield provides only an upper limit (Table 1), corresponding
to a rate of 3.5 counts/kg/d in the {$^{40}$K} peak, which is
consistent with the measured one (Fig. 3-a).

\subsubsection{Summary}
Radioactivity measurements and associated simulations have
shown that most of EDELWEISS-I gamma background arises from a tiny
U/Th contamination of the very massive copper shield rather than
from radioactive contaminations close to the detectors.

\section{Backgrounds from alpha particles and surface events}
\label{par:alpha}

\subsection{Alpha particles}
One of the most noticeable features revealed in the 2003h data is
the presence of a distinct population localized at a recoil energy
$E_R=5.33\pm 0.03$ MeV with a quenching factor \footnote{The
quenching factor is defined as the relative ionization efficiency
between a nuclear recoil and an electron recoil of the same real
recoil energy.}  $Q=0.30\pm 0.02$, consistent over the three
detectors (see Fig. 4). This population was later confirmed in the
2003p data using the high-energy reconstruction procedure detailed
in section 3.2 (see Ref.~\cite{FioTh}). The rates vary from $2.4\pm
0.6$ to $5.0\pm 0.8$ counts/kg/d in the fiducial volumes 
(center electrodes) of the three detectors, and from $13\pm 2$ to
$25\pm 2$ counts/kg/d in the lateral volumes (guard
electrodes). We observe a significantly higher rate in the top
detector than in the bottom detector (Table 2), a fact for
which we have no explanation.

Given the energy range and peculiar value of the quenching factor,
we link this population to the interaction of alpha particles in our
detectors. Our explanation is that the detectors themselves and/or
their close environment, i.e. their copper holders, suffer from a
$^{210}$Pb contamination. This isotope, with a half-life of 22.3
years, is a daughter of $^{222}$Rn. It can be implanted on a surface
exposed to an atmosphere containing radon during the fabrication and
handling of the detectors. The last disintegration of the chain
$^{210}$Pb$\rightarrow ^{210}$Bi$\rightarrow ^{210}$Po$\rightarrow
^{206}$Pb produces an alpha with an energy of 5.3 MeV. If the events
with $Q<0.5$ on Fig. 4 are interpreted as alphas, their energy
spectrum restricts their origin to either the detectors themselves
or the copper surfaces ($\leq 1\mu$m) surrounding them. Indeed,
given the very low penetration length of alpha particles of such
energies in germanium or copper ($\sim10\mu$m, see Table 3), a
volume contamination of any material other than the germanium itself
would undoubtedly lead to the observation of an alpha energy
continuum down to 0~MeV. While we do observe such a tail to low
energies, it is clearly not the dominant feature.

The localization in quenching is also interesting. Although such a
phenomenon may be associated with incomplete charge collection due
to surface interactions, it is difficult to explain why we observe a
constant value of $Q\sim0.3$, and not a range of values down from
$Q=1$. Previous studies~\cite{Qalpha,Klap} have shown that charge
collection efficiency for alpha interactions in Ge is similar to
that of gammas. However, those measurements were obtained at room or
liquid nitrogen temperatures, and under a drifting field of several
thousands V/cm. Despite the lack of more relevant results, it is not
unreasonable to assume that, in our case, because of the particular
nature of an alpha interaction in germanium and the high local
density of charges created, some systematic recombination before
collection takes place. This would typically lead to a constant
value of the quenching ratio, as is observed in our data.

\subsection{Heavy nuclear recoils}
When  $^{210}$Po decays to $^{206}$Pb a 5.3 MeV alpha is produced
and the $^{206}$Pb nucleus recoils with a kinetic energy of 103~keV.
Depending on whether this takes place at the surface of the detector
itself or at the surface of the copper in front of the germanium,
what we expect to see is substantially different. In both cases, the
penetration length of such a heavy particle at such a low energy is
so small (a few tens of nm) that all of the ionization signal is
lost, charges being absorbed either in the aluminum electrode
(100~nm) or in the amorphous sub-layer (another 60~nm) if the Pb
nucleus hits a surface not covered by an electrode. If the
contamination is localized on the surface of the detector, the heat
signal should correspond to the full recoil energy of 103~keV. In
contrast, if the Pb nucleus originates from the copper, then it has
to go through a thickness of material equivalent to its implantation
depth before it reaches the detector, resulting in a partial loss of
energy.

In order to isolate this heavy recoil population, we looked at all
the events in our data compatible with a signal above detection
threshold on the heat channel, and below threshold on the ionization
channel. This led to a classification into three categories: a)
sensor "NTD" events, induced inside the small thermal sensor by its
own radioactivity; b) random noise events, linked for example to
anomalous microphonic episodes inside the cryostat, and c)
"legitimate" ionization-less events, among which potentially lie the
heavy recoils we seek. As described in Ref.~\cite{San05}, NTD events
present a very particular pulse shape, significantly shorter than
regular bulk events. We use shape discrimination to eliminate
populations (a) and (b) and isolate population (c) (see
Ref.~\cite{FioTh} for details).

Using the data from run 2003p, in each detector we find a population
of events clearly contained below 100 keV (Fig. 5), with rates
varying from $1.5\pm 0.5$ to $5.4\pm 0.8$ counts/kg/d in the total
volume of the detectors for recoil energies greater than 40~keV (see
Table 2). Interestingly, as was the case with alpha interactions,
the top detector registers more events than the bottom one, and in
the same ratio within the error limits. This tends to confirm that
both populations are indeed linked. The fact that we do not see a
clear peak at 103~keV but a roughly uniform front below this energy
also implies that the contamination is localized exclusively on the
copper surfaces, and not on the detectors themselves.

\subsection{Surface beta interactions}

If the contamination is indeed linked to an exposure to radon, then
we expect to see the products of all disintegrations in the
$^{210}$Pb chain, in addition to the 5.3 MeV alphas and heavy
recoils. In particular, the $^{210}$Pb decay to $^{210}$Bi leads to
a complex spectrum of low energy conversion and Auger electrons
together with two beta spectra with end-points at 63.5 keV and
17 keV. Due to the value of its penetration length in
germanium (see Table 3), the 46.5~keV gamma ray does not contribute
significantly to the surface event budget. The decay of $^{210}$Bi
to $^{210}$Po emits another beta electron (branching fraction 100\%)
with an end-point of 1.16 MeV.

According to Monte-Carlo simulations using the CASINO
code~\cite{Casino} (see Table 3), an electron of 100 keV impinging
at normal incidence the germanium surface will lose 90\% of its
energy in the first 20 $\mu$m. This value goes up to $\sim$700
$\mu$m for a 1 MeV electron. As for the nuclear recoils (sect. 4.2),
the 100 nm aluminum electrode and the 60 nm amorphous semi-conductor
sub-layer constitute a "dead ionization zone" in our detectors. As
for the alphas (sect. 4.1), there exists a zone under the electrodes
where electron-hole pairs are not properly collected. The depth of
this zone can be as large as 10 $\mu$m, as studies using detectors
with a different design have shown~\cite{Shu00,cdms}.

We therefore expect to observe some incomplete charge collection for
a significant part of the events generated by the decay electrons.
In the 2003p data, we do indeed observe a population of
"intermediate" events between the electron recoil band and the
nuclear recoil band (Fig. 6). As shown in Ref.~\cite{San05}, this
population is absent when the detectors are exposed to a $^{137}$Cs
source of penetrating gamma rays. In order to quantify this
population, we compare our low-background data to $^{137}$Cs gamma
calibration data. This allows us to define an area in the $(Q, E_R)$
plane (with $E_R\le$80 keV) where we can be confident that events
are not caused by gamma or neutron interactions with full charge
collection~\cite{FioTh}. This selection underestimates the actual
total population of incomplete charge collection events. In order to
provide an order-of-magnitude estimate of the rate of surface betas,
we assume a 50\% selection efficiency as hinted by Fig. 13 of
Ref.~\cite{San05}. The results are given in Table 2: the count rates
are similar to those of alpha particles. Furthermore, the same
ratios between the counting rates in each detector are observed. We
also notice that the intermediate population appears to display
energies below $\sim$ 60~keV, which is the maximum energy for an
electron from a $^{210}$Pb beta decay. These observations point
toward a correlation between the identified alpha population and
this intermediate population.

\subsection{Contamination scenario}

In conclusion, we have identified three event populations consistent
with a single coherent contamination scenario:
\begin{itemize}
\item A population of alpha interaction events from the disintegration of
    $^{210}$Po very close to the detectors, with rates of the order of 5 counts/kg/d
    in the fiducial volumes of the detectors.
\item A population of ionization-less events, with energies below 100 keV, which
    we can identify with $^{206}$Pb recoils originating from the copper surfaces
    facing the detectors. The count rates are of the same order as those observed
    for alpha interactions, and the rate ratios between detectors are consistent with alpha data.
\item A population of events with ionization yields lying between those for gamma
    rays and those for nuclear recoils, associated to surface interactions, which we
    link to electrons emitted in the decay of $^{210}$Pb. Here again, the rates are of the
    same order as those observed for alpha interactions, and the ratios between detectors are conserved.
\end{itemize}

This leads us to strongly suspect the $^{210}$Pb contamination of
the copper parts facing the detectors while the hypothesis of a
contamination of our detectors themselves is not favoured due to the
absence of registered ionization-less events with a nominal energy
of 103~keV.

It is worth mentioning that the gamma component of the background
produces also Compton electrons which can escape from the copper
surfaces surrounding the detectors. Miscollected surface events can
be produced when such low energy electrons hit a detector. The
simulation of the dominant gamma background (from the bulk of the
copper shield) has shown that, below 200 keV, 0.6~\% (1.1~\%) of the
events are electrons interacting in the first 20~$\mu$m (50~$\mu$m)
under the electrodes; electrons coming from outside of the detectors
constitute 70~\% of these populations. The absolute rates are
between 2 counts/kg/d (20~$\mu$m) and 4 counts/kg/d (50~$\mu$m), the
actual fraction of miscollected electrons depending on the variation
of the charge collection efficiency with depth.

Finally, a complete model of beta contamination on the detector and
copper surfaces should also consider $^{14}$C. All surfaces are
usually quickly covered by a thin (about 1.5-4 nm) layer of organic
compounds CH$_x$ (see for instance~\cite{Drag}). Given an isotopic
ratio $^{14}$C/$^{12}$C$=1.3\times 10^{-12}$, this yields 0.5 to 3
beta electrons/kg/d for the central part of a detector, with a 156 keV
end-point. Comparison with ``intermediate event'' rates given in
Table 2 shows that there might be a contribution from $^{14}$C
contamination.

\section{Neutron background}
\label{par:neutron}

Fast (0.5-10 MeV) neutron interactions are a dangerous background as
they produce nuclear recoils that can mimic a WIMP signal. One
coincidence between nuclear recoils in two detectors due to a
neutron was observed in EDELWEISS-I data ~\cite{San05}. We present
in this section the simulations of this neutron background.

\subsection{Simulation of neutron transport in EDELWEISS-I}
\label{par:neutron-cal} Before any attempt to evaluate the number of
nuclear recoils in the EDELWEISS~-~I data due to interactions from a
residual neutron flux, it is necessary to test the reliability of
the Monte Carlo code used for neutron transport through a simulation
of a calibration run with a $^{252}$Cf source. Using GEANT3 a
normalization factor of 0.68 had to be applied to the simulation to
reproduce the experimental rates~\cite{phdsanglard}. An alternative
simulation was then  performed with MCNPX~\cite{MCNP}. The detector
response was applied to the simulated germanium recoils as described
in Ref.~\cite{martineau}. The data and the simulation were then
analyzed using the same selection procedure to extract the recoil
energy spectra.

The $^{252}$Cf source was positioned, through a 50~mm diameter
hole in the lead/copper shield, at 70~mm from the external wall of
the lead shield and at 20mm below the detector stack: the neutron
count rate decreased from the bottom (GGA3) to the top (GSA3)
detector. These rates are well reproduced by the simulations as can
be seen in Fig.~\ref{fig:n-cal}. The ratio MCNPX/data of total
number of events is 1.067 $\pm$ 0.079 (measured activity of the
source~\cite{phdsanglard}) $\pm$ 0.015 (Stat) $\pm$ 0.007 (MC Stat).
The measured proportion of events in the 3 detectors of 27\%, 32\% and 40\% is well reproduced by the simulation (30\%, 32\% and 40\%). However
the shapes of the energy spectra differ slightly, a feature also
observed with GEANT3 simulations.

\subsection{Neutron flux in the Modane underground laboratory}
\label{par:neutron-lsm}

With a rock overburden equivalent to 4800~m of water, the residual
neutron background originates mainly from the radioactivity of the
rock. Two contributions arise from the contamination of U/Th:
spontaneous fission and ($\alpha$,n) reactions.In Ref.~\cite{Cha96},
the neutron flux coming out of the LSM rock had been deduced from
the measurements made with a $^6$Li-doped liquid scintillator cell,
using GEANT3 for the neutron transport simulation. This spectrum was
in turn interpreted in terms of the sum of the contributions from
spontaneous fission and ($\alpha$,n) reactions, using a simplified
calculation for the latter process. In Ref.~\cite{neutrons}, the
original data had been re-interpreted in the light of more reliable
neutron transportation codes (optimized versions of GEANT3 and
MCNPX), having for consequence the reduction of the estimated
neutron flux in the LSM from 4.0~$10^{-6}$ to
1.6~$10^{-6}$~n/cm$^2$/s. In the present work (first introduced in
Ref.~\cite{lem06}), we conclude these studies by using the
SOURCES~\cite{wils} code, which includes a more exact calculation of
the ($\alpha$,n) contribution and using MCNPX for neutron transport.

The neutron flux in the LSM was simulated using the procedure
described in~\cite{lem}. The SOURCES code was used to calculate the
rate and energy spectrum of neutrons produced by spontaneous fission
and ($\alpha$,n) reactions due to the U/Th contamination in the
rock. The LSM rock composition and its U/Th contamination was taken
from Ref.~\cite{Cha96}.  The neutron propagation in the rock to the
LSM cavity takes into account backscattering on the walls. In the
present calculation, the concrete covering the walls is not present.

The measurement of Ref.~\cite{Cha96} was performed with a
$^6$Li-doped NE320 liquid scintillator cell,  where neutrons were
positively identified from the observation of a proton recoil
followed by the neutron absorption on $^6$Li. The detector
(8.5$\times$8.5$\times$85 cm$^3$ scintillator) and its shielding (5
cm Cu and 12 cm Pb) were simulated with the MCNPX code.  The
detection efficiency is about 10$\%$ for incident neutrons above an
energy threshold of 2~MeV. The following treatments were applied:
light yield efficiency based on Birks' law fitted to proton data,
10$\%$ energy resolution, 43$\%$ event selection
efficiency~\cite{Cha96}.

Fig.~\ref{fig:n-meas} shows the comparison of the simulations to the
data. The SOURCES spectrum agrees better in shape with the data than
the spectrum determined in Ref.~\cite{Cha96}.  Note however that all
simulations are normalized to data. A normalization factor of 2.28
is needed in the simulation, which could be explained by
inhomogeneities in the rock composition of the LSM walls, especially
its water content. The large influence over the neutron flux of a
small amount of hydrogen (mainly in water) in the LSM rock is
emphasized in Ref.~\cite{lem}: the 1\% hydrogen content reduces the
neutron flux above 1 MeV by a factor 2.1 (see also Ref.~\cite{Wulan}
for a discussion of this effect in the Gran Sasso context).

Fig.~\ref{fig:n-lsm} shows the results of the simulation for the
neutron spectrum in the LSM cavity after normalization to the data.
We take the normalization factor as a systematic uncertainty on the
simulation. We obtain a flux of
1.06~$\pm$0.10(stat.)~$\pm$0.59(syst.)~$10^{-6}$~n/cm$^2$/s above
1~MeV. Assuming a full coverage of the walls by concrete and its
U/Th contaminations taken from Ref.~\cite{Cha96} would give
essentially the same result on the flux (less than 1$\%$ higher)
with a corresponding normalization factor of 2.03.

\subsection{Neutrons in EDELWEISS-I}
\label{par:neutron-edw}

The rate of nuclear recoils due to the neutron flux coming out of
the rock, as estimated in Fig.~\ref{fig:n-lsm}, has been calculated
using MCNPX. This flux is transported through the 30 cm paraffin
shielding and the experimental setup shown in Fig.~1.

We expect about 0.026~$\pm$0.002(MC stat)~$\pm$0.018(syst)
neutrons/kg/d from the rock radioactivity, 0.002 $\pm$1.2$\%$ (MC
stat) neutrons/kg/d from the 0.25 ppb $^{238}$U contamination of
copper shield and less than 0.001 neutrons/kg/d from the upper 0.1
ppb limit on $^{238}$U contamination in lead shield. This translates
into about 1.6~$\pm$0.1(MC stat)~$\pm$1.1(syst)  nuclear recoils
expected in EDELWEISS~-~I data (62 kg$\cdot$day).
Fig.~\ref{fig:n-edw} shows the corresponding recoil spectra in the
detectors. The experimental spectrum of Ref.~\cite{San05}, with 34
events between 15 and 30 keV and 3 betwen 30 and 100 keV, can easily
accommodate the presence of a few nuclear recoils due to neutron
scattering. The only direct experimental proof of the presence of a
neutron flux is the observation of a coincidence between nuclear
recoils in two detectors~\cite{San05}. The present simulation gives
further support for this interpretation. First, the rock spectrum of
Fig.~\ref{fig:n-edw} shows that the recoil energies in this
coincidence (14.8 and 14.5 keV) are typical for neutron scattering.
Secondly, the simulation confirms that the ratio of single to
coincidence is approximately 10:1, consistent with the ratio
observed in neutron calibrations. With this ratio, it was concluded
in Ref.~\cite{San05} that from one coincidence, the prediction range
for accompanying single events is from zero to 40 events at
90$\%$~CL. Although consistent with the hypothesis of a background
originating from both surface events and neutron interactions, this
range does not help constrain the relative importance of these two
contributions.

\section{Conclusions and prospects}
\label{par:concl}

Three main backgrounds have been identified in the EDELWEISS-I data
sets. The sensitivity is limited by background events which, after
the nuclear recoil selection, mimic true WIMP induced nuclear
recoils with a low value, about 0.3, of the ionization to recoil
signals ratio $Q$. EDELWEISS-II, the second phase of the experiment,
is designed for a two orders of magnitude sensitivity improvement
relying upon the efficient rejection of these background events.

Before the rejection of the bulk electron recoils, the overwhelming
majority of the events is a gamma population entirely dominated by
the U/Th contamination of the Cu shield. These events are rejected
with an efficiency greater than 99.9\%. Nevertheless Compton
electrons can escape from the closest copper surfaces, reach the
detectors and possibly produce miscollected surface events; the
gamma component has then to be maintained as low as possible. In the
EDELWEISS-II set up, extensive material selection, mounting and
operation of the cryostat under clean room conditions (class 100),
secure a better radioactive cleanliness. Copper is no longer used
for shielding. The inner part of the shield is made of very low
radioactivity archeological lead. Based on GEANT3 simulations and the results of the activity measurements, the overall gamma rate for this new setup is predicted to be between 1.0 and 0.1 of the gamma rate of EDELWEISS-I, using upper limits or central values, respectively.

A very small $^{210}$Pb contamination at the surface of the copper
detector casings is the very likely source of the observed alphas
and near-electrode electron events showing a deficit of the charge
collection. This probably comes from an exposure to radon at some
steps of the manufacturing and handling process. In order to
minimize the radon exposure, the EDELWEISS-II clean room is
supplied, during the detector mounting phase, with air of very low
radon concentration delivered by the LSM radon-trap facility.
Nevertheless a more decisive approach consists in the identification
and rejection of the near-surface events. Detectors equipped with
Nb$_x$Si$_{1-x}$ thin films as thermal sensors have been operated in
the last months of the EDELWEISS-I phase. These films are sensitive
to the transitory athermal part of the phonon signal, which
constitutes a near-surface tag~\cite{CSNSM}. These first tests with
200 g modules have shown a factor of ten reduction of the surface
event rate while retaining a 80\% efficiency. Seven 400g NbSi
modules, over a total of 28 detectors, will be operated in the first
phase of EDELWEISS-II. Other possible solutions are still in a R\&D
phase: identification of surface events using interdigitized
electrodes or pulse shape analysis of the charge
signal~\cite{Bronia}.

The last identified background component comes from neutrons, which,
as WIMPs, induce nuclear recoils through elastic scattering. The
corresponding count rate is not yet much constrained by EDELWEISS-I,
but its existence is established by the observation of one
coincidence event. For EDELWEISS-II the reduction of this neutron
background becomes critical. A 50 cm thick polyethylene shield all
around the experiment moderates the low energy neutrons. Neutrons
created by muon interactions in the shielding are tagged with a 5 cm
thick plastic scintillator muon veto of 100 m$^2$ surrounding the
whole experiment (95\% coverage). The compact arrangement of the
multi (up to one hundred) detector structure allows further
rejection through anti-coincidence between detectors.

EDELWEISS-II is now running at the Laboratoire Souterrain de
Modane and first results are scheduled for 2007.

\mbox{}\\\noindent{\large\bf Acknowledgments }\\
The help of the technical staff of the Laboratoire Souterrain de
Modane and of the participant laboratories is gratefully
acknowledged. This work has been partially supported by the EEC
Applied Cryodetector network (Contracts ERBFMRXCT980167 and
HPRN-CT-2002-00322) and the ILIAS integrating activity (Contract
RII3-CT-2004-506222).

\newpage


\begin{table*}
\begin{tabular}{| c | c | c || c | c | c| }
\hline
& U & Th & $^{40}$K & $^{60}$Co & $^{210}$Pb \\
& (ppb) & (ppb) & (mBq/kg) & (mBq/kg) & (mBq/kg) \\
\hline
\hline
Copper shield & $0.25\pm0.06$ & $0.44\pm0.27$ & $<15$ & $<0.6$ & $300\pm150$\\
Copper holders & $<0.1$ & $<0.1$ & & $<1.0$ & \\
Cub1 springs & $<0.6$ & $<1.7$ & $40\pm30$ & & $260\pm130$ \\
Roman lead & $<0.022$ & $<0.032$ & $<1.3$ & & $<200$ \\
Teflon balls & $<0.2$ & $<0.5$ & $80\pm30$ & $5\pm2$ & $40\pm20$ \\
Wires & & & $1400\pm1000$ & &\\
\hline
\end{tabular}
\caption{\label{Table 1} Measured contaminations or activities for several materials used in EDELWEISS-I. Statistical errors are at 1 $\sigma$ and limits at 90\% confidence level (the detector background at 46.5 keV leads to high errors on $^{210}$Pb activities).}
\end{table*}

\begin{table*}

\begin{tabular}{| c | c  c | c  c | c  c | }
\hline
Detector & \multicolumn{2}{|c|}{GSA3} & \multicolumn{2}{|c|}{GSA1} & \multicolumn{2}{|c|}{GGA3} \\
Electrode & Center & Guard & Center & Guard & Center & Guard \\
\hline
\hline
Alphas count rate (/kg/d) & 5.0$\pm$0.8 & 24.7$\pm$2.2 & 5.2$\pm$0.8 & 17.8$\pm$1.7 & 2.4$\pm$0.6 & 13.3$\pm$1.5 \\
\hline
\hline
Heavy nuclear recoils & &&&&&\\
Count rate (/kg/d)&\multicolumn{2}{|c|}{$5.4\pm0.8$} & \multicolumn{2}{|c|}{$2.1\pm0.6$} & \multicolumn{2}{|c|}{$1.5\pm0.5$} \\
\hline
\hline
"Intermediate events" & &&&&&\\
 Count rate (/kg/d) & 6.3$\pm$1.0 & 33.0$\pm$2.5 & 5.0$\pm$0.8 & 33.3$\pm$2.4 & 3.2$\pm$0.7 & 20.6$\pm$1.9 \\
\hline
\end{tabular}
\caption{\label{Table 2} Count rates for alpha events, heavy nuclear recoil events (recoil energy greater than 40 keV) and "intermediate events" (between 15 and 200~keV recoil energy). These latter are defined as the events between the electron recoil band at 3.29$\sigma$ and the nuclear recoil band at 1.65$\sigma$ (see Fig. 6).}
\end{table*}

\begin{table*}
\begin{tabular}{| l  c | c | c | c |}
\hline
Particle & Energy & Cu & Ge & Pb \\
\hline
\hline
      & 10 keV  & 9 $\mu$m & 170 $\mu$m & 18 $\mu$m  \\
Gamma & 100 keV & 6 mm     & 8 mm       & 400 $\mu$m \\
      & 1 MeV   & 40 mm    & 80 mm      & 30 mm      \\
\hline
         & 10 keV  & 200 nm     & 350 nm     & \\
Electron & 100 keV & 11 $\mu$m  & 20 $\mu$m  & \\
         & 1 MeV   & 340 $\mu$m & 700 $\mu$m & \\
\hline
Alpha & 5.3 MeV & 11 $\mu$m & 19 $\mu$m & 15 $\mu$m \\
\hline
Polonium & 100 keV & 40 nm & 68 nm & \\
\hline
\end{tabular}
\caption{\label{Table 3} Typical penetration lengths of various particles in Cu, Ge and Pb. Values for gammas correspond to a 10~\% transmission probability. Values for electrons correspond to the maximum depth at which an electron has deposited 90~\% of its energy (results from CASINO~\cite{Casino} simulations). Values for alphas and Po nuclei are the mean penetration length given by SRIM~\cite{SRIM} simulations with a normal incidence.}
\end{table*}


\begin{figure}
\includegraphics[width=1.0\textwidth]{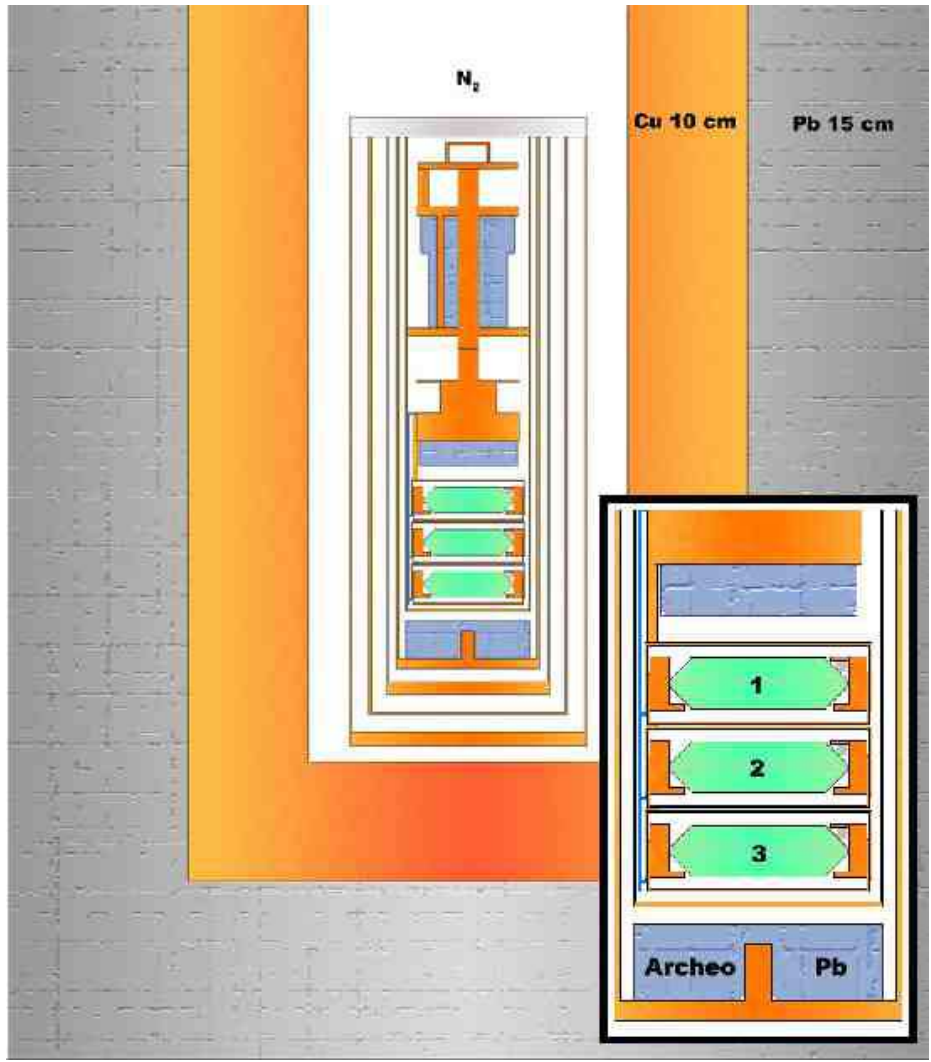}
\caption{Schematic view of the EDELWEISS-I cryostat within its Cu and Pb shields,
    as implemented in GEANT3 Monte Carlo simulations. Orange areas figure copper,
    grey textured areas stand for lead. The inset shows the three germanium detectors
    (from top to bottom: GSA3, GSA1 and GGA3) encased into individual copper casings. Teflon balls and Cu springs are visible in the upper right corner of the detectors holders. Wires to the cold electronics are going up at the left of the detector stack.
    Not represented on the figure is the 30~cm external paraffin shield against neutrons.}
\label{setup}
\end{figure}

\begin{figure}
\includegraphics[width=1.0\textwidth]{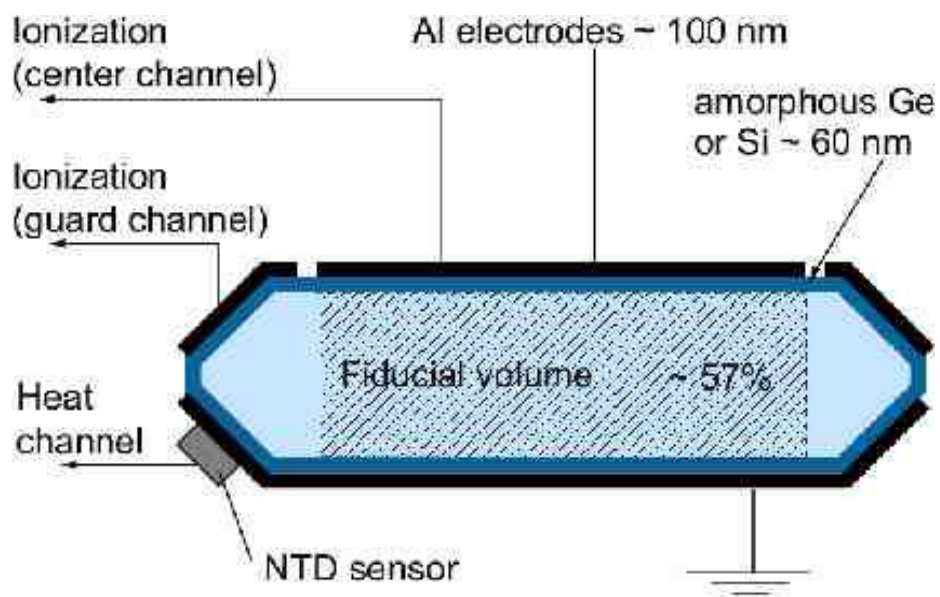}
\caption{Schematic cut of an EDELWEISS heat and ionization germanium detector.
    The thickness of Al electrodes, amorphous layer and NTD sensor are not represented to scale.}
\label{setup-det}
\end{figure}

\begin{figure}
\includegraphics[width=1.2\textwidth]{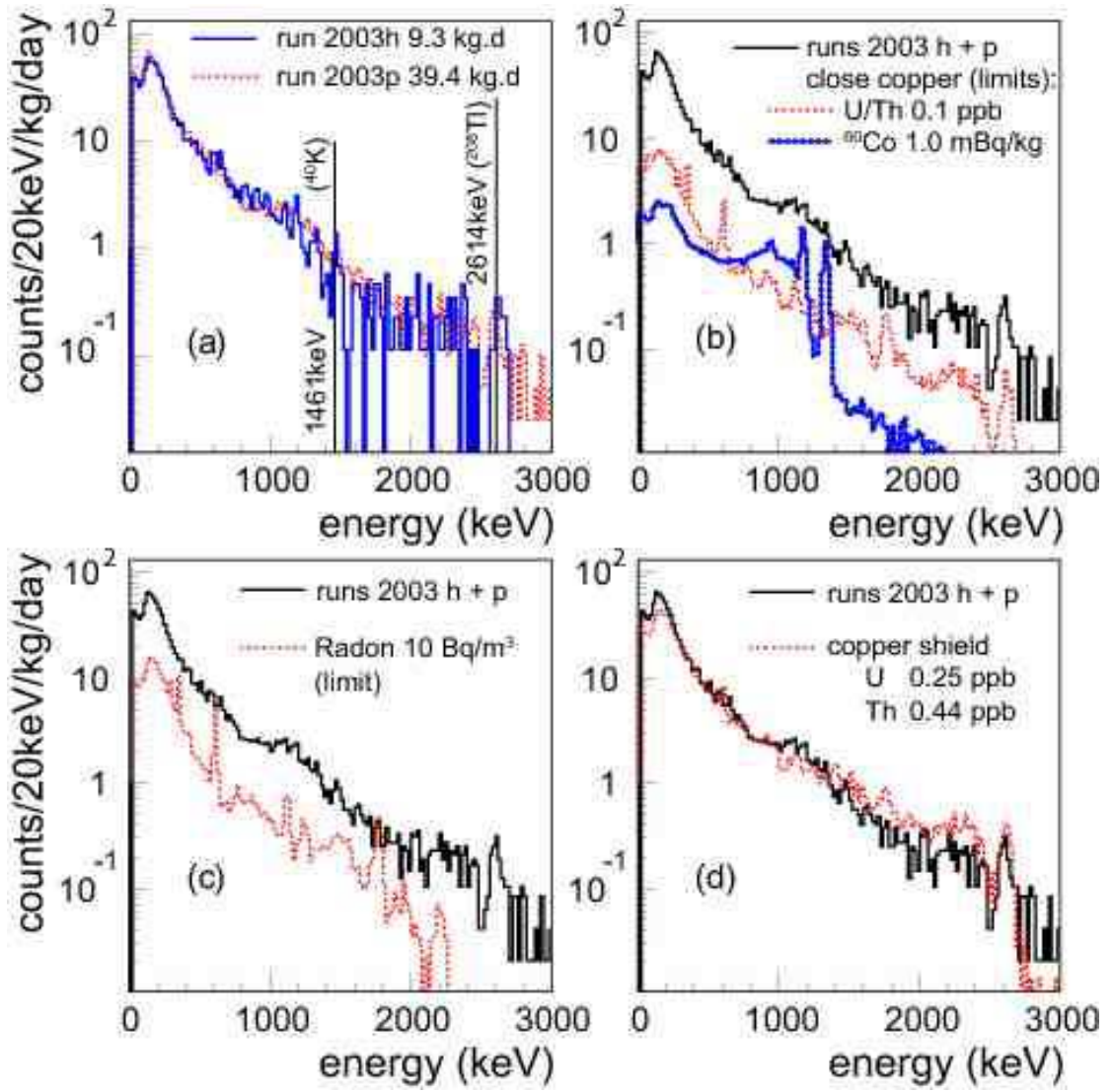}
\caption{Gamma background energy spectrum (ionization channel) compared to Monte
    Carlo simulations of various sources of radioactive contamination. (a) The two experimental
    data sets 2003h (full line) and 2003p (dashed line). (b) Sum of the 2003p and 2003h data sets
    (full line), simulations of the U/Th (dashed line) and $^{60}$Co (dotted line) contents of
    copper close to the detectors (detector holders, cryostat). The quoted radioactive contaminations
    are measured upper limits. (c) Sum of the 2003p and 2003h data sets (full line), simulated
    contribution of the radon trapped in the lead-copper shield (upper limit, dotted line).
    (d) Sum of the 2003p and 2003h data sets (full line), simulation of the U/Th content of
    the copper shield (measured concentrations, dashed line).}
\label{spec1b}
\end{figure}

\begin{figure}
\includegraphics[width=1.0\textwidth]{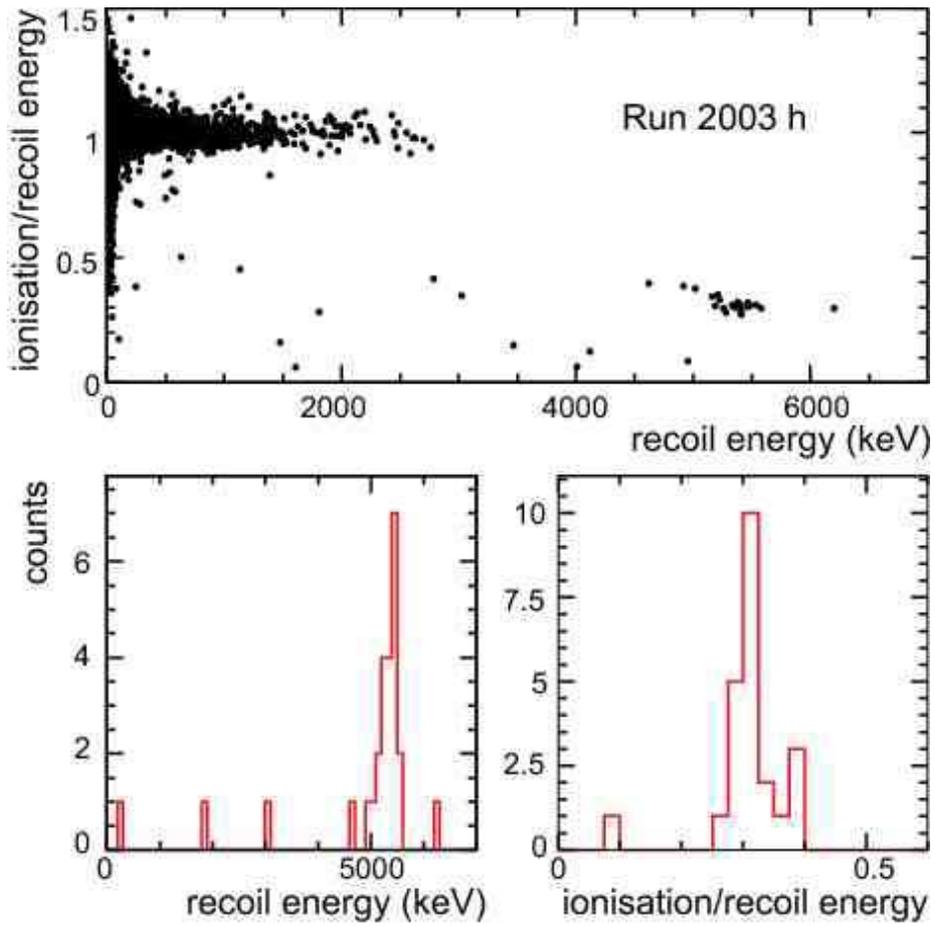}
\caption{Distribution of the ionization quenching ratio $Q$ with recoil energy $E_R$, for the
    data of the run 2003h. Bottom left: $E_R$ projection for $0.2<Q<0.4$.\break Bottom right: $Q$
    projection for $4.5<E_R<6$ MeV.}
\label{fig:qalph}
\end{figure}

\begin{figure}
\includegraphics[width=38pc]{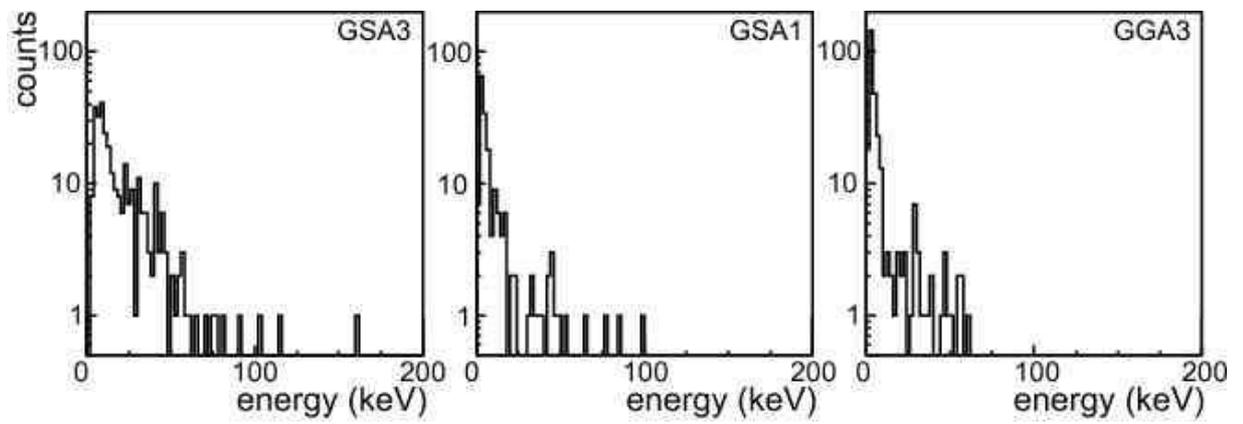}
\caption{Amplitude distributions of ionization-less events after rejection of NTD events and
    noise events. The heavy nuclear recoil events at the surfaces belong to this population.}
\label{fig:ioless}
\end{figure}

\begin{figure}
\includegraphics[width=1.0\textwidth]{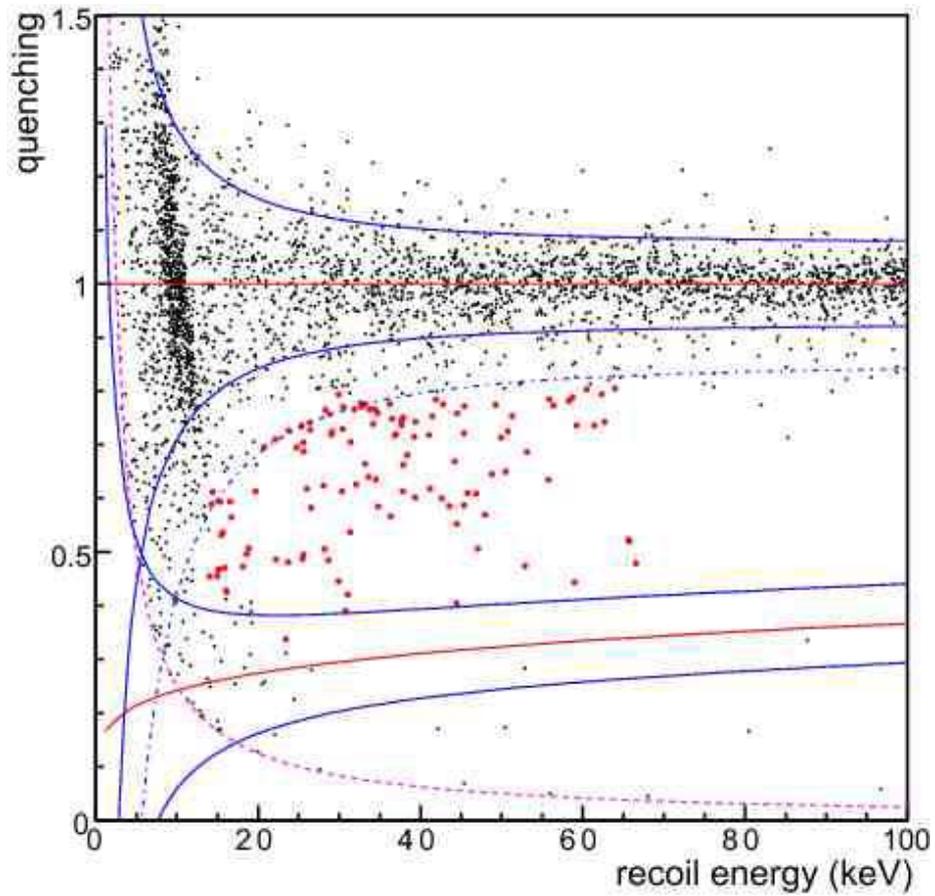}
\caption{Representative sample of the ``intermediate event'' population: run 2003p, sum of
    the fiducial volumes of the three detectors, 22.5 kg.d. Full (red) circles: selected
    ``intermediate events''; (black) dots: remaining part of the low background data.}
\label{fig:intermediate}
\end{figure}

\begin{figure}[h]
\begin{center}
\includegraphics[width=30pc]{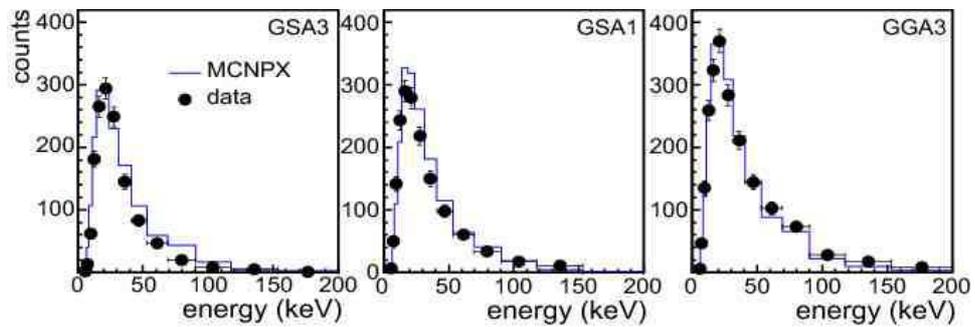}
\caption{\label{fig:n-cal} Energy spectra of the nuclear recoils in EDELWEISS-I calibration
    runs with a $^{252}$Cf neutron source. The MCNPX simulation is also shown (full line,
    no normalization to data).}
\end{center}
\end{figure}

\begin{figure}[h]
\begin{minipage}{15pc}
\includegraphics[width=15pc]{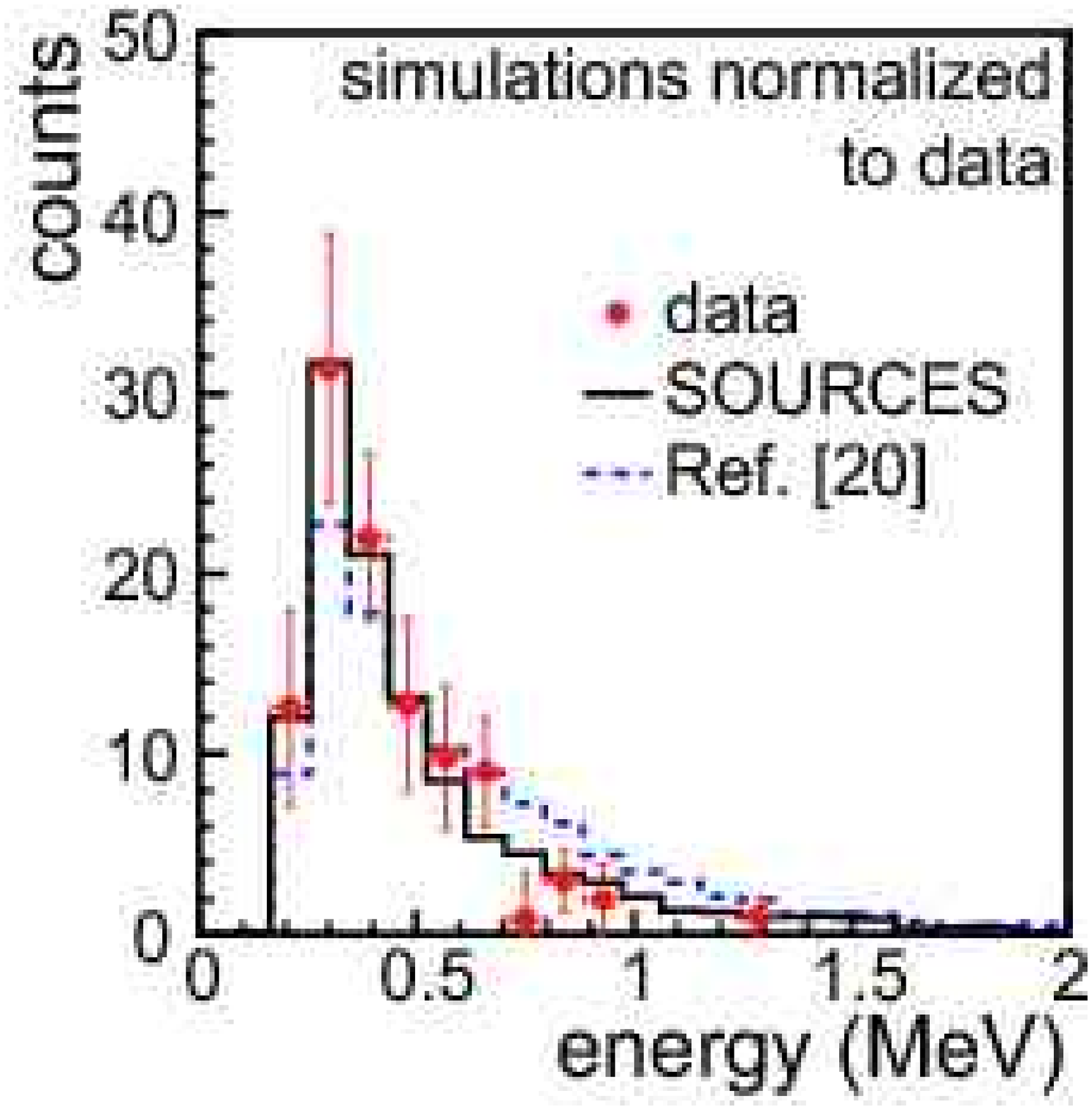}
\caption{\label{fig:n-meas} Neutron flux measurements in the LSM cavity from Ref.~\cite{Cha96}.
    The data points are the experimental electron equivalent energy spectrum. The dotted line is
    the original simulated spectrum from rock radioactivity~\cite{Cha96}. The full line corresponds
    to the present work using SOURCES and MCNPX. Both simulations are normalized to the data.}
\end{minipage}\hspace{2pc}%
\begin{minipage}{15pc}
\includegraphics[width=15pc]{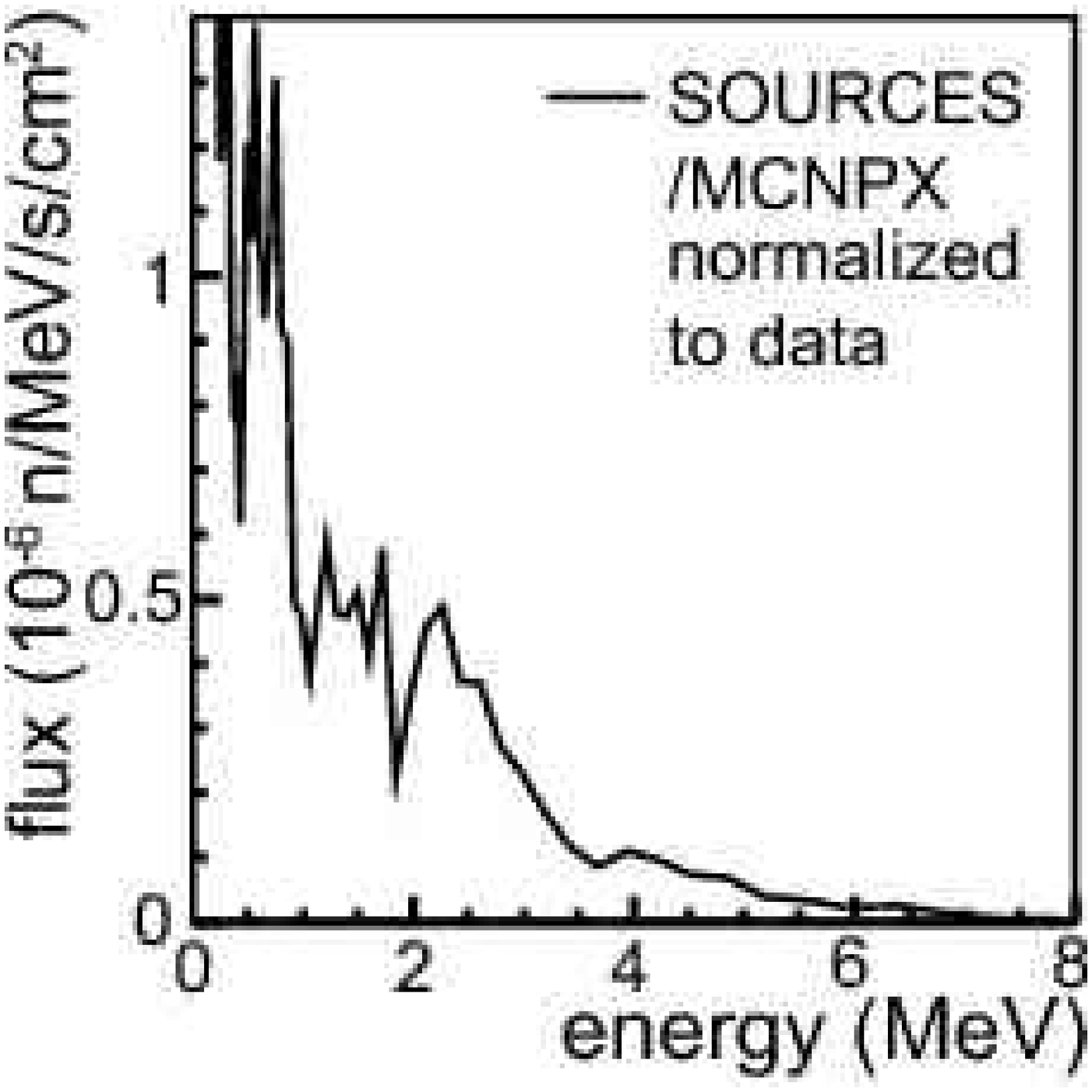}
\caption{\label{fig:n-lsm}Simulated neutron energy spectrum in the LSM after normalization
    to data of Ref.~\cite{Cha96}. The neutron production from $^{238}$U and $^{232}$Th traces in
    the rock is simulated with SOURCES and propagated with MCNPX.}
\end{minipage}
\end{figure}

\begin{figure}[h]
\begin{center}
\includegraphics[width=18pc]{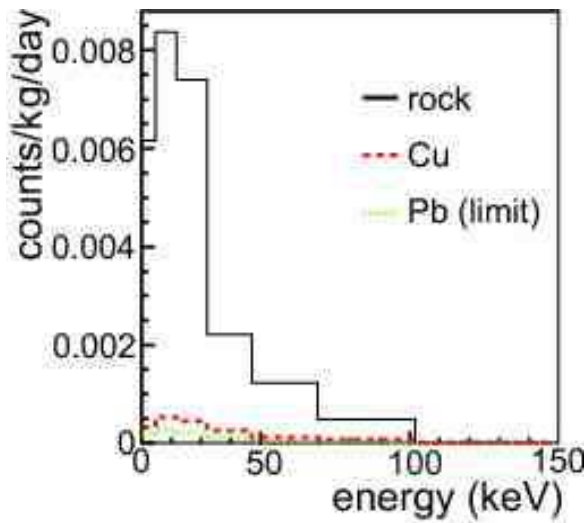}
\caption{\label{fig:n-edw} Expected energy spectra of the nuclear recoils in EDELWEISS-I
    low-background  runs from radioactivity of the rock (full line), $^{238}$U contamination
    in copper (dashed line) and in lead (dotted line).}
\end{center}
\end{figure}

\end{document}